\documentclass[a4paper, 11pt]{article}
\pdfoutput=1

\usepackage{jcappub}

\usepackage[T1]{fontenc}
\usepackage{graphicx}
\usepackage{float}
\usepackage{placeins}
\usepackage{tablefootnote}
\usepackage{pbox}
\usepackage{booktabs}
\usepackage{makecell}
\usepackage{multirow}
\usepackage{array}
\usepackage{mleftright}
\usepackage[ISO]{diffcoeff}
\usepackage{csquotes}
\usepackage{siunitx}
\usepackage{hyperref}
\usepackage[capitalise]{cleveref}

\makeatletter
\input{aas_macros.sty}
\let\jnl@style=\relax
\makeatother

\DeclareSIUnit{\pc}{pc}
\DeclareSIUnit{\kpc}{\kilo\pc}
\DeclareSIUnit{\Mpc}{\mega\pc}
\DeclareSIUnit{\Gpc}{\giga\pc}
\DeclareSIUnit{\hHubble}{\text{\ensuremath{h}}}
\crefname{section}{\S}{\S}
\crefformat{section}{\S#2#1#3}
\crefmultiformat{section}{\S#2#1#3}{ and~\S#2#1#3}{, \S#2#1#3}{ and~\S#2#1#3}
\AddToHook{cmd/appendix/before}{%
    \crefalias{chapter}{appendix}%
    \crefalias{section}{appendix}%
    \crefalias{subsection}{appendix}%
}

\title{Testing the 4-Point Correlation Function as a Probe of Parity Violation with the \textnormal{\textsc{Quijote-Odd}} Simulations}

\author[1]{William Ortolá Leonard,}
\author[2]{Zachary Slepian,}
\author[3]{Alex Krolewski,}
\author[4]{Simon May}

\affiliation[1]{Department of Physics, University of Florida,\\2001 Museum Rd., Gainesville, FL 32611, USA}
\affiliation[2]{Department of Astronomy, University of Florida,\\211 Bryant Space Science Center, Gainesville, FL 32611, USA}
\affiliation[3]{California Institute of Technology,\\1200~East California Boulevard, Pasadena, CA 91125, USA}
\affiliation[4]{Fakultät für Physik, Universität Bielefeld,\\Universitätsstraße 25, 33615 Bielefeld, Germany}

\emailAdd{wortola@ufl.edu}
\emailAdd{zslepian@ufl.edu}
\emailAdd{akrolews@caltech.edu}
\emailAdd{simon.may@uni-bielefeld.de}

\abstract{We investigate parity violation searches with the 4-Point Correlation Function (4PCF) using the \textsc{Quijote} and \textsc{Quijote-Odd} halo catalogs. Specifically, we explore the auto- and cross-correlation searching methodologies that have been previously used in the Baryon Oscillation Spectroscopic Survey (BOSS) and the Dark Energy Spectroscopic Instrument (DESI) searches. Although the \textsc{Quijote-Odd} simulations are initialized with a parity-violating primordial trispectrum, we find no statistically detectable parity-odd signal in the halo 4PCF. Instead, we find that the halo measurements are dominated by Poisson noise. We are able to distinguish this noise-dominated behavior from the response expected for a parity-odd signal. These results clarify how parity-sensitive 4PCF statistics respond to noise in a controlled simulation setting.}

\begin{document}

\maketitle

\section{Introduction}
Parity is a fundamental spatial symmetry describing the invariance of a physical system under the inversion of its spatial coordinates, $\mathbf{x}\rightarrow-\mathbf{x}$. In a 3-dimensional (3D) setting, the 4-Point Correlation Function (4PCF) is the lowest-order statistic capable of detecting whether the Universe is parity-invariant (PI) or parity-violating (PV) \cite{cahn_parity, shiraishii, jeong_fossil}. A pair of points contains no notion of handedness, while three points define a triangle that, in 3D, can be mapped onto its mirror image by a rotation.\footnote{In 2D, the 3PCF is the corresponding lowest-order correlation function capable of detecting PV signals in the Universe as demonstrated in \cite{Ortola_2d_Test_PV}.} Four points, however, define a tetrahedron and provide three independent separation vectors, from which one may construct the scalar triple product $\mathbf{r}_1\cdot(\mathbf{r}_2\times\mathbf{r}_3)$ \cite{cahn_iso, cahn_parity}. This quantity is invariant under rotations but changes sign under a parity transformation, allowing tetrahedra of opposite handedness to be distinguished. Consequently, the 4PCF can be decomposed into parity-even and parity-odd components, with the expectation value of the latter vanishing for a statistically PI Universe. This makes the parity-odd 4PCF a particularly direct probe of Parity Violation (PV). 

The first searches for PV using the galaxy 4PCF were performed with the Sloan Digital Sky Survey (SDSS) Baryon Oscillation Spectroscopic Survey (BOSS) \cite{hou_parity, phil_parity}, where measurements of the parity-odd 4PCF initially produced intriguing evidence for a departure from the parity-symmetric expectation. Ref.~\cite{hou_parity}, using the BOSS LOWZ and CMASS samples, reported significances of approximately $3.1\sigma$ and $7.1\sigma$, respectively. Ref.~\cite{phil_parity}, using the CMASS sample and a non-parametric analysis, found a smaller but still notable significance of approximately $2.9\sigma$. Subsequent work, \cite{Krolewski_No_PV_det}, demonstrated that such measurements are particularly sensitive to any possible data–mock mismatch for the construction of the covariance. In particular, \cite{Krolewski_No_PV_det} reanalyzed BOSS and found a mismatch reduced the inferred significance to between $0$ and $2.5\sigma$. 

More recently, measurements with the Dark Energy Spectroscopic Instrument (DESI) Year~1 Luminous Red Galaxy (LRG) sample have highlighted a closely related issue \cite{Slepian_Parity_Meas_DESI}. An auto-correlation analysis---detailed more fully in \cref{sec:statistics_methodology}---found large apparent excesses relative to the mock catalogs, reaching roughly $4$--$10\sigma$. On the contrary, a cross-correlations analysis---also explained in more detail in \cref{sec:statistics_methodology}---between independent regions of the sky did not detect a corresponding parity-odd signal. A separate DESI analysis, \cite{Hou_Parity_odd_2}, also found a PV signal when using the auto-correlation analysis, but concluded that the results were consistent with zero once a possible mismatch between the data and mocks was taken into account. 

The current observational situation emphasizes that a robust search for PV requires a detailed understanding of how the adopted covariance, mock catalogs, data vector compression, and test statistic behave, both when parity is absent, and when the underlying initial conditions contain PV physics. The \textsc{Quijote} and \textsc{Quijote-Odd} halo simulations in principle should provide such a controlled environment in which these questions can be investigated \cite{villaescusa2020quijote, coulton}. The former supplies parity-conserving realizations, while the latter is initialized with a parity-violating primordial trispectrum. However, we found that this primordial PV does not produce a statistically detectable parity-odd signal in the halo 4PCF considered here. Consequently, these simulations cannot provide a complete validation of the sensitivity of current 4PCF methodologies to a genuine halo-level parity-violating signal. However, we found that the 4PCF statistics can distinguish correlated noise from a genuine parity-odd contribution.

The structure of this work is as follows: In \cref{Sec:Quij_data}, we present the \textsc{Quijote} and \textsc{Quijote-Odd} simulations. In \cref{sec:statistics_methodology}, we explain the methodology to search for PV 4PCF signals. Then, we present the results of this work in \cref{sec:Results} and finish with discussion and conclusions in \cref{Sec:Discussion}.

\section{\textnormal{\textsc{Quijote}} Simulations: Full Region \& Sub-regions}
\label{Sec:Quij_data}

The \textsc{Quijote} simulations consist of 500 cosmological $N$-body simulations, designed to study the statistical properties of the large-scale structure (LSS) of the Universe and its dependence on cosmological parameters \cite{villaescusa2020quijote}. The simulations evolve dark matter particles in periodic boxes of side length \qty{1}{\per\hHubble\Gpc}, with the \textsc{Quijote-Odd} extension introducing PV directly into the initial conditions \cite{coulton}. Specifically, the initial conditions are modified such that the primordial trispectrum contains a parity-odd amplitude, controlled by the parameter $p_{\mathrm{NL}}$. The \textsc{Quijote-Odd} suite contains two sets of 500 realizations, denoted ODD$_{\mathrm{p}}$ and ODD$_{\mathrm{m}}$, with $p_{\mathrm{NL}}=+10^{6}$ and $p_{\mathrm{NL}}=-10^{6}$, respectively. Importantly, realization $i$ of ODD$_{\mathrm{p}}$, ODD$_{\mathrm{m}}$, and the corresponding fiducial simulation are generated from the same underlying Gaussian random field. Thus, the three simulations share the same complex phases, while differing in the PV contribution in their initial conditions. This matched construction is particularly useful for separating effects associated with the injected parity violation from those arising from cosmic variance, as explained in \cref{sec:Results}.

We perform our analysis on the halo catalogs from the full simulation volumes, using \num{500} ODD$_{\mathrm{p}}$, \num{500} ODD$_{\mathrm{m}}$, and 500 Fiducial realizations, for a total of \num{1500} full-box measurements. These we term the Full Regions (FR). In addition, we extract two non-overlapping sub-regions from each realization in order to investigate the behavior of the 4PCF as the analyzed volume is reduced and perform a cross-correlation analysis (explained in more detail in the paragraph around \cref{eq:chi2_cross_quijote}). Sub-region A (Sub~A) is a cube of side length \qty{400}{\per\hHubble\Mpc}, with a volume of \qty{0.064}{\per\hHubble\cubed\Gpc\cubed}, while Sub-region B (Sub~B) is a cube of side length \qty{300}{\per\hHubble\Mpc} and a volume of \qty{0.027}{\per\hHubble\cubed\Gpc\cubed}. The sub-regions are positioned such that they are separated from one another by \qty{100}{\per\hHubble\Mpc}. Additionally, we require a \qty{100}{\per\hHubble\Mpc} separation between each sub-region and the nearest boundary of the \qty{1}{\per\hHubble\Gpc} simulation box. This requirement prevents the FR periodic wrapping from making distant halos artificially close and keeps the two sub-regions clearly separated as independent volumes.

\section{Methodology}
\label{sec:statistics_methodology}

Here, we briefly outline our methodology to measure a PV signal with the 4PCF. We note that all measurements of the 4PCF described below are using \textsc{CADENZA}---a GPU version of the code \textsc{ENCORE} \cite{encore}. The estimator proceeds by selecting each halo in turn as a primary and expanding the surrounding galaxy distribution into spherical harmonics on a set of spherical, concentrical shells \cite{iso_gen, cahn_iso}. Products of the resulting harmonic coefficients are then combined and averaged over all primaries to obtain the 4PCF coefficients. In this work, we use 18 linearly spaced radial bins spanning $r_{\mathrm{min}} = \qty{20}{\per\hHubble\Mpc}$ to $r_{\mathrm{max}} = \qty{160}{\per\hHubble\Mpc}$. This corresponds to a radial bin width of
\begin{equation}
    \Delta r = \frac{r_{\mathrm{max}}-r_{\mathrm{min}}}{18}
    \simeq \qty{7.8}{\per\hHubble\Mpc},
\end{equation}
which also sets the smallest radial scale interval entering the tetrahedral configurations used in our measurements. More details on the estimator and algorithm can be found in \cite{Ortola_4PCF_Meas_DESIY1, phil_4pcf, hou_parity}.

We quantify the parity-odd 4PCF measurements using two complementary statistics. The first, which we denote by $T^2$, uses a covariance estimated from the simulations. The second, denoted by $\chi^2$, uses the analytical covariance of \cite{hou_covar}, and the algorithm of \cite{Analytic4PC, Krolewski_No_PV_det}. The distinction between the two approaches therefore lies primarily in the covariance and whether the 4PCF data vector is first compressed to a restricted set of modes (see next paragraph for a more detailed explanation).

For the $T^2$ analysis, we begin from the Gaussian analytical covariance of the 4PCF, $\mathbf{C}_{\mathrm{Ana.}}$ \cite{hou_covar}, and perform its eigendecomposition,
\begin{equation}
    \mathbf{C}_{\mathrm{Ana.}}
    =
    \mathbf{E}\boldsymbol{\Lambda}\mathbf{E}^{\mathsf{T}},
\end{equation}
where $\boldsymbol{\Lambda}$ is diagonal and contains the eigenvalues, and the columns of $\mathbf{E}$ are the corresponding eigenvectors. The analytical covariance is used at this stage only to define a low-noise basis in which the 4PCF signal can be compressed. For each mode, the 4PCF signal is projected onto the corresponding eigenvector and divided by the variance of that mode. Schematically, the signal-to-noise contribution of eigenmode $\alpha$ is
\begin{equation}
    \left(\frac{S}{N}\right)_{\alpha}
    =
    \frac{\left|\mathbf{e}_{\alpha}^{\mathsf{T}}\boldsymbol{\zeta}\right|}{\sqrt{\lambda_{\alpha}}},
    \label{eq:eigenmode_sn}
\end{equation}
where $\lambda_{\alpha}$ and $\mathbf{e}_{\alpha}$ are respectively the eigenvalue and eigenvector of mode $\alpha$. $\boldsymbol{\zeta}$ denotes the parity-odd 4PCF data vector. The eigenmodes are then ordered according to their signal-to-noise (S/N) contribution.

After this ordering, we retain only the first $N_{\mathrm{eig.}}$ highest S/N modes, and project each 4PCF measurement into the corresponding reduced-dimensional space,
\begin{equation}
    \widetilde{\boldsymbol{\zeta}}
    =
    \mathbf{E}_{N_{\mathrm{eig.}}}^{\mathsf{T}}
    \boldsymbol{\zeta},
    \label{eq:projected_zeta}
\end{equation}
where $\mathbf{E}_{N_{\mathrm{eig.}}}$ is the set of chosen eigenvectors. Throughout this work, we consider
\begin{equation}
    N_{\mathrm{eig.}} = \numlist{50; 100; 300},
\end{equation}
allowing us to test the stability of the results as progressively more information from the original 4PCF data vector is retained.

The covariance entering $T^2$ is then determined empirically from the projected simulation measurements rather than from the analytical covariance itself. Importantly, when evaluating $T^2$ for realization $i$, that realization itself is excluded from the covariance estimate. This \enquote{leave-one-out empirical} covariance is defined as
\begin{equation}
    \widehat{\mathbf{C}}^{i}_{\mathrm{Emp.}} = \frac{1}{N_{\mathrm{s}}-2} \sum_{j\neq i} \left(\widetilde{\boldsymbol{\zeta}}_{j} - \overline{\widetilde{\boldsymbol{\zeta}}}\right) \times \left(\widetilde{\boldsymbol{\zeta}}_{j}-\overline{\widetilde{\boldsymbol{\zeta}}}\right)^{\mathsf{T}},
    \label{eq:loo_empirical_cov}
\end{equation}
where $N_{\mathrm{s}}$ is the total number of realizations used and $i$ is the excluded realization. The corresponding test statistic is
\begin{equation}
    T_i^2
    =
    \widetilde{\boldsymbol{\zeta}}_{i}^{\mathsf{T}}
    \left(
        \widehat{\mathbf{C}}^{i}_{\mathrm{Emp.}}
    \right)^{-1}
    \widetilde{\boldsymbol{\zeta}}_{i}.
    \label{eq:T2_auto_def}
\end{equation}
This leave-one-out procedure ensures that the realization being tested does not contribute to the empirical covariance against which it is compared. The analytical covariance therefore determines which modes are retained, while the simulations themselves determine the covariance of those modes entering the final $T^2$ calculation. Further details concerning the covariance treatment and eigenmode compression procedure are in Appendix~B of \cite{Ortola_4PCF_Meas_DESIY1}. 

Our second method, $\chi^2$, only uses the analytical covariance \cite{hou_covar, Analytic4PC}. In this case, no empirical covariance is constructed from the simulations, and we instead define
\begin{equation}
    \chi^2
    =
    \boldsymbol{\zeta}^{\mathsf{T}}
    \mathbf{C}_{\mathrm{Ana.}}^{-1}
    \boldsymbol{\zeta}.
    \label{eq:chi2_auto_def}
\end{equation}
We term the two analyses described above the \enquote{auto-correlation analyses} since they will use a data vector from the same region of the simulation. 

In addition to these auto-correlation analyses, we construct cross-correlation analyses. The cross-correlation computes test statistics for the 4PCF measurements between two different regions of the simulations.\footnote{More than two regions can be formed, but for the cross-correlation analyses of this work we restrict ourselves to two: Sub A $\times$ Sub B, and ODD$_{\mathrm{p}}$ $\times$ ODD$_{\mathrm{m}}$.} In general, for two data vectors $\boldsymbol{\zeta}_{a}$ and $\boldsymbol{\zeta}_{b}$ of different sub-regions, the corresponding analytical covariance-based test statistic is
\begin{equation}
    \chi^2_{\times}
    =
    \boldsymbol{\zeta}_{a}^{\mathsf{T}}
    \mathbf{C}_{\mathrm{Ana.}}^{-1}
    \boldsymbol{\zeta}_{b}.
    \label{eq:chi2_cross_quijote}
\end{equation}
An analogous quantity can be defined in the compressed space for the $T^2$ analysis using the leave-one-out empirical covariance. In contrast to the auto-correlation, the cross-correlation is not necessarily positive and may take either positive or negative values. A positive value indicates that the two 4PCF data vectors are preferentially aligned (\textit{i.\,e.,} either both are greater than zero or both are less than zero), whereas a negative value indicates anti-alignment between the two 4PCF vectors.

This distinction is particularly useful for the matched-phase ODD$_{\mathrm{p}}$ and ODD$_{\mathrm{m}}$ simulations. Computing the cross-correlation between ODD$_{\mathrm{p}}$ and ODD$_{\mathrm{m}}$ should result in a negative-value-centered distribution: The cross-correlation analysis discards uncorrelated noise and isolates the PV signal \cite{Krolewski_No_PV_det}. We therefore expect that the cross-correlation between ODD$_{\mathrm{p}}$ and ODD$_{\mathrm{m}}$ results in a negative-centered distribution as a result of the different injected PV signals, $p_{\mathrm{NL}}= \pm 10^6$. This motivates a cross-correlation analysis between matched-phase realizations of ODD$_{\mathrm{p}}$ and ODD$_{\mathrm{m}}$. 

\begin{figure}
    \centering
    \includegraphics[width=0.5\textwidth]{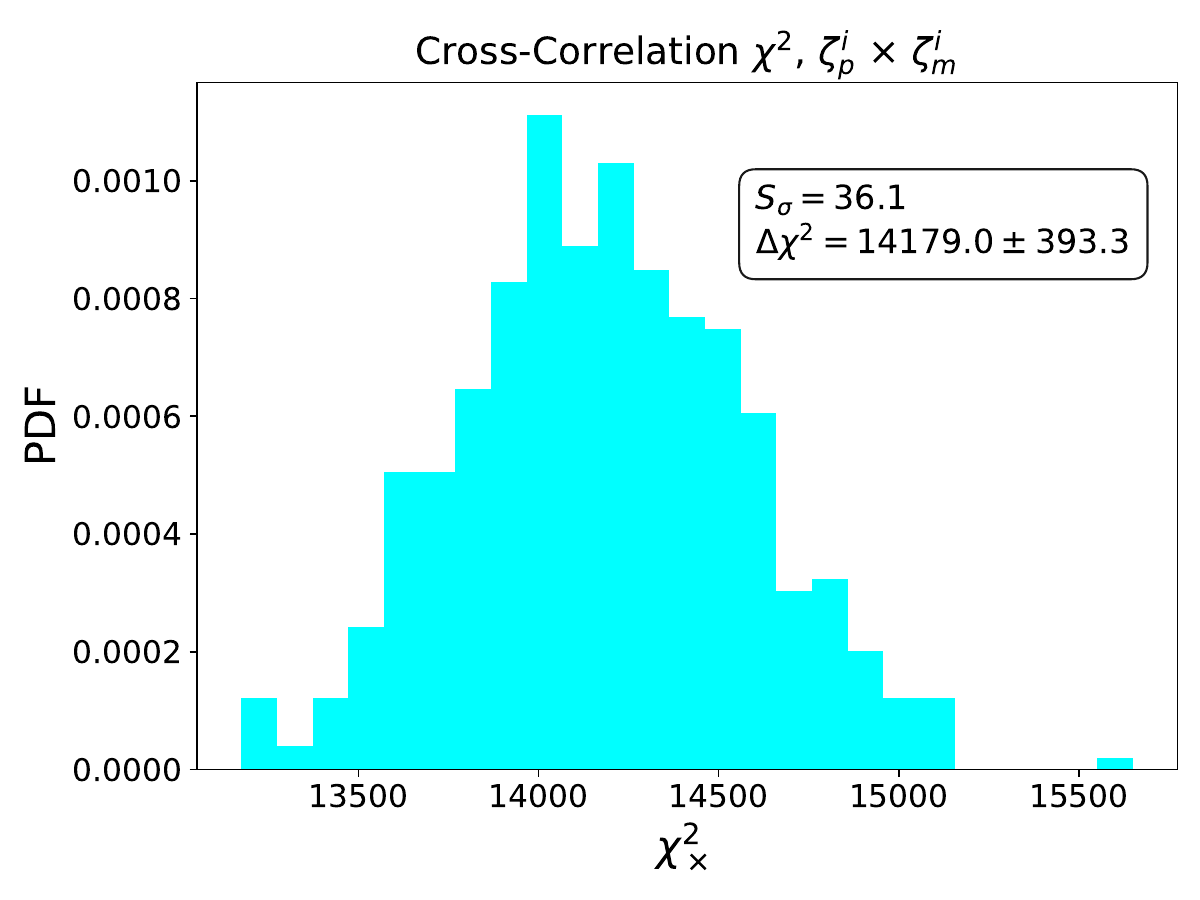}%
    \includegraphics[width=0.5\textwidth]{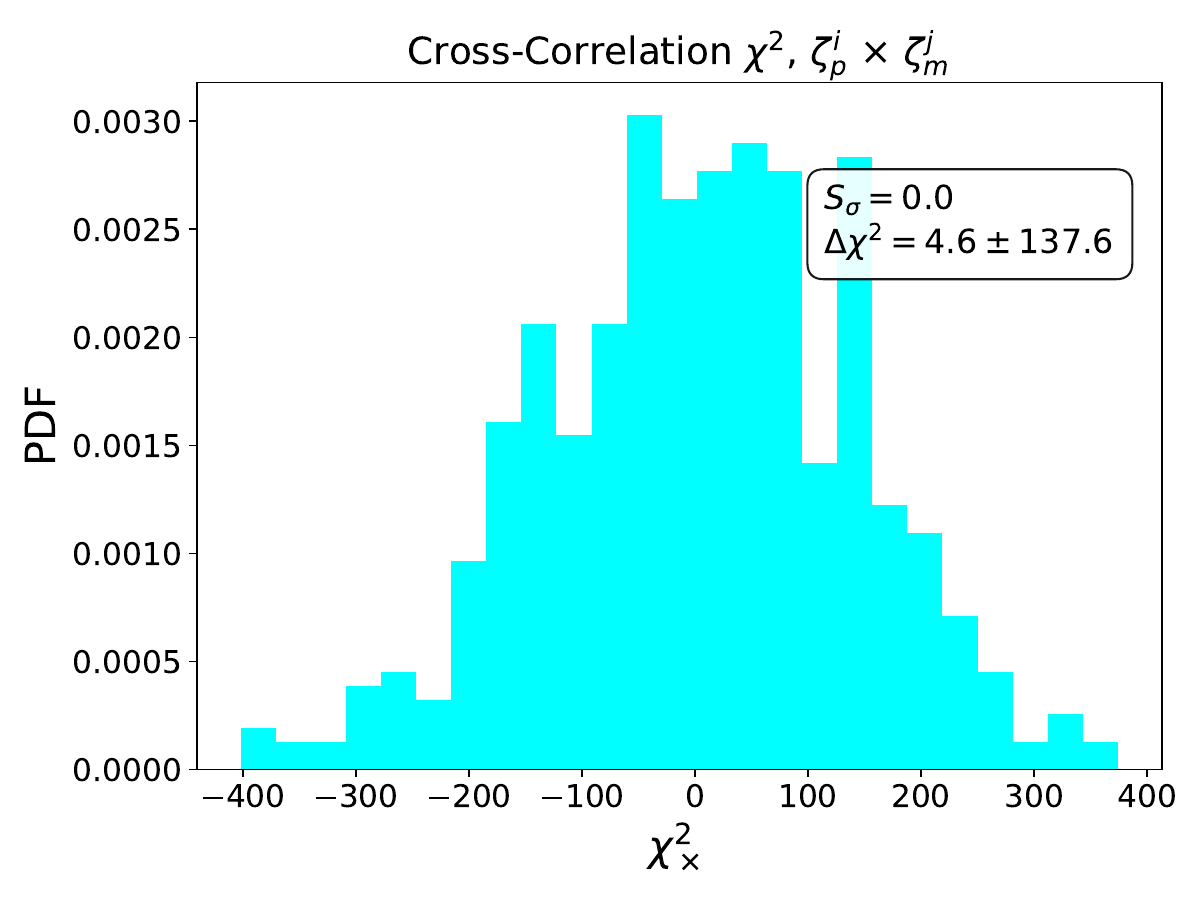}
    \caption{Cross-correlation $\chi^2_{\times}$ distributions for the ODD$_{\mathrm{p}}$ and ODD$_{\mathrm{m}}$ halo catalogs. The left panel shows the result obtained by cross-correlating phase-matched realizations, where ODD$_{\mathrm{p}}$ and ODD$_{\mathrm{m}}$ share the same initial phases. The right panel shows the corresponding out-of-phase analysis obtained by pairing different realization numbers. The phase-matched case produces a strong positive detection of $36.1\sigma$, whereas the signal disappears completely in the out-of-phase case, yielding $0.0\sigma$. This behavior shows that the large $\text{ODD}_{\mathrm{p}} \times \text{ODD}_{\mathrm{m}}$ correlation is driven by the noise shared between simulations with the same complex phases and does not persist when this common contribution is removed.}
    \label{fig:Phase_&_outofPhase_ODDp_X_ODDm}
\end{figure}

However, as shown in the left panel of \cref{fig:Phase_&_outofPhase_ODDp_X_ODDm}, the matched-phase cross-correlation results in a positive value, contrary to expectation. This further motivates the out-of-phase analysis in the right panel of \cref{fig:Phase_&_outofPhase_ODDp_X_ODDm}. The results demonstrate that out-of-phase realizations have uncorrelated noise and that \textsc{Quijote-Odd} contains no PV signal at the halo level. We explain this in more detail in \cref{sec:Results,Sec:Discussion}.

\begin{figure}
    \centering
    \includegraphics[width=0.5\textwidth]{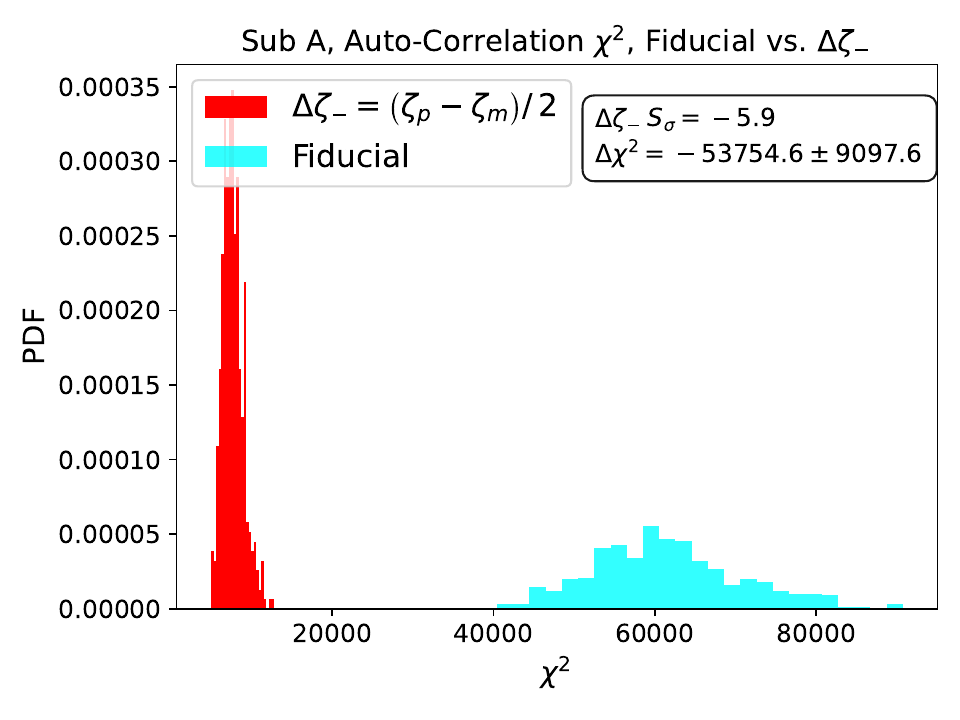}%
    \includegraphics[width=0.5\textwidth]{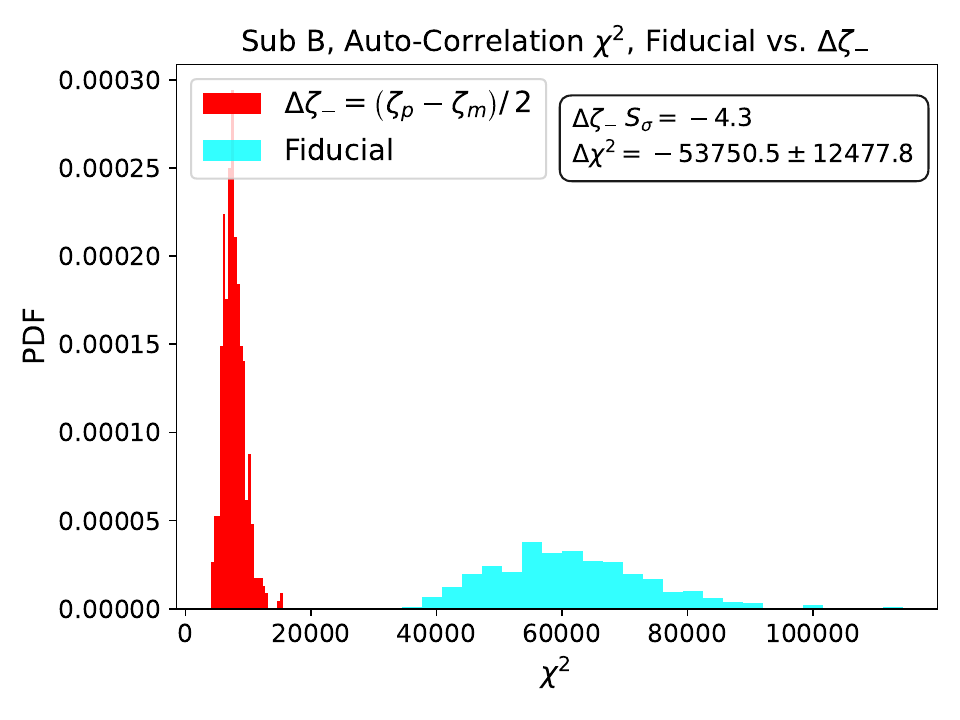}
    \caption{Auto-correlation $\chi^2$ distributions for $\Delta\boldsymbol{\zeta}_{-}$ measured in Sub~A and Sub~B. The left panel shows the comparison between the Fiducial and $\Delta\boldsymbol{\zeta}_{-}$ distributions in Sub~A, while the right panel shows the corresponding result for Sub~B. In both cases, the $\Delta\boldsymbol{\zeta}_{-}$ distribution lies well below the Fiducial distribution, with significances (in terms of $\sigma$) of $S_{\sigma}=-5.9\sigma$ and $S_{\sigma}=-4.3\sigma$ for Sub~A and Sub~B, respectively. The same behavior is  observed in the FR analysis, demonstrating that the $\text{ODD}_{\mathrm{p}}-\text{ODD}_{\mathrm{m}}$ subtraction continues to remove a large fraction of the shared noise while leaving a residual contribution.}
    \label{fig:auto_chi2_subA_subB_zeta_minus}
\end{figure}

Finally, the opposite signs of the parity-violating amplitudes in ODD$_{\mathrm{p}}$ and ODD$_{\mathrm{m}}$ motivate a difference-vector analysis:
\begin{equation}
    \Delta\boldsymbol{\zeta}_{-}
    =
    \frac{1}{2}
    \left(\boldsymbol{\zeta}_{\mathrm{p}}-\boldsymbol{\zeta}_{\mathrm{m}}\right),
    \label{eq:zeta_minus_quijote}
\end{equation}
and a sum-vector analysis:
\begin{equation}
    \Delta\boldsymbol{\zeta}_{+}
    =
    \frac{1}{2}
    \left(\boldsymbol{\zeta}_{\mathrm{p}}+\boldsymbol{\zeta}_{\mathrm{m}}\right).
    \label{eq:zeta_plus_quijote}
\end{equation}
For matched realizations, contributions such as shot noise, cosmic variance, \textit{etc.}, common to ODD$_{\mathrm{p}}$ and ODD$_{\mathrm{m}}$ should be suppressed by the difference-vector analysis. On the contrary the PV signal---which reverses sign between the two simulations---should add up in the difference-vector analysis. The exact opposite should occur in the sum-vector analysis. We note that $\Delta\boldsymbol{\zeta}_{-}$ and $\Delta\boldsymbol{\zeta}_{+}$ can be analyzed with either of the covariance treatments described above. Results for these analyses can be found in \cref{fig:subtraction_and_addition_FR_chi2_result,fig:auto_chi2_subA_subB_zeta_minus,tab:auto_results,tab:cross_results}.

\section{Results}
\label{sec:Results}

We present the auto- and cross-correlation analyses results for the Fiducial, ODD$_{\mathrm{p}}$, and ODD$_{\mathrm{m}}$ halo catalogs, as well as for the combinations $\Delta\boldsymbol{\zeta}_{-}$ and $\Delta\boldsymbol{\zeta}_{+}$ defined in \cref{eq:zeta_minus_quijote,eq:zeta_plus_quijote}. These tests are carried out for both $\chi^2$ and $T^2$, and for the full simulation boxes and the two sub-regions. In this section, we show a subset of the results that most clearly illustrates the main conclusions of the analysis. The complete set of measurements is made available through Zenodo \cite{supplementary_images_Quijote_Halos_PV}.

\begin{figure}
    \centering
    \includegraphics[width=0.5\textwidth]{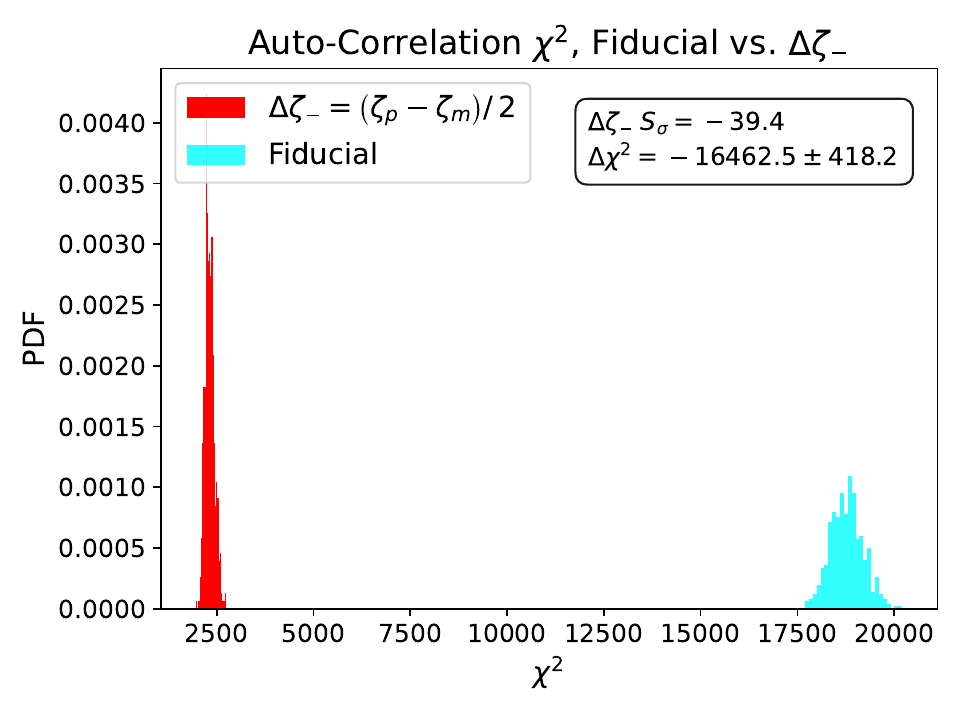}%
    \includegraphics[width=0.5\textwidth]{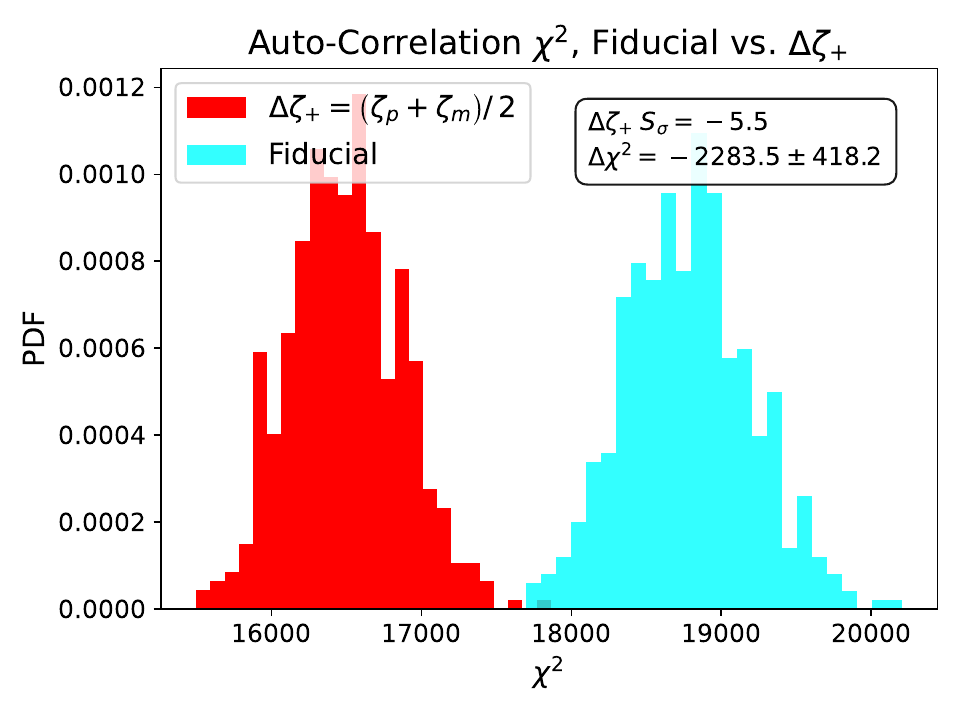}
    \caption{Auto-correlation $\chi^2$ distributions for the full \textsc{Quijote} simulation volume. The left panel compares the Fiducial distribution with $\Delta\boldsymbol{\zeta}_{-}$, while the right panel compares the Fiducial distribution with $\Delta\boldsymbol{\zeta}_{+}$. For $\Delta\boldsymbol{\zeta}_{-}$, the distribution is strongly shifted below the Fiducial case, corresponding to a significance (in terms of $\sigma$) of $S_{\sigma}=-39.4\sigma$, whereas $\Delta\boldsymbol{\zeta}_{+}$ shows a smaller discrepancy of $S_{\sigma}=-5.5\sigma$. The smaller values of $\Delta\boldsymbol{\zeta}_{-}$ with respect to the fiducial indicates that the subtraction of the phase-matched ODD$_{\mathrm{p}}$ and ODD$_{\mathrm{m}}$ realizations removes a large fraction of their shared noise, while no detectable parity-odd signal is recovered in the halo catalogs.}
    \label{fig:subtraction_and_addition_FR_chi2_result}
\end{figure}

We first consider the auto-correlation measurements of $\Delta\boldsymbol{\zeta}_{-}$ and $\Delta\boldsymbol{\zeta}_{+}$, with the results shown in \cref{fig:subtraction_and_addition_FR_chi2_result} and summarized in \Cref{tab:auto_results}. Opposite to expectation, $\Delta\boldsymbol{\zeta}_{-}$ lies far below the Fiducial distribution, with a difference of $-39.4\sigma$ in the $\chi^2$ analysis. As previously discussed in \cref{sec:statistics_methodology}, $\Delta\boldsymbol{\zeta}_{-}$ is constructed to isolate the PV signal (assuming the signal is sufficiently strong); we would therefore expect its distribution to lie above the Fiducial distribution. Instead, we find a strong suppression, providing the first indication that no detectable PV signal is present in the \textsc{Quijote-Odd} halo 4PCF. On the contrary, $\Delta\boldsymbol{\zeta}_{+}$ should suppress the antisymmetric PV contribution and more closely resemble the Fiducial distribution, yet we find a discrepancy of $-5.5\sigma$ in the $\chi^2$ statistic. Taken together, these results indicate that the noise present in the \textsc{Quijote-Odd} halo catalogs is not statistically identical to that of the Fiducial simulations, while the $\text{ODD}_{\mathrm{p}} - \text{ODD}_{\mathrm{m}}$ subtraction suppresses much of the common contribution and leaves only a residual noise component. We find the same behavior when analyzing Sub~A and Sub~B. As explained below, the cross-correlation analysis provides an independent test of this interpretation.

When ODD$_{\mathrm{p}}$ and ODD$_{\mathrm{m}}$ realizations with the same realization number are cross-cor\-re\-lat\-ed, we find a strong positive correlation, reaching $36.1\sigma$ in the $\chi^2_{\times}$ analysis and $5.3\sigma$ in the corresponding $T^2_{\times}$ analysis (for $N_{\mathrm{eig.}}=100$). Since the matched ODD$_{\mathrm{p}}$ and ODD$_{\mathrm{m}}$ simulations are generated from the same underlying Gaussian realization, this result shows that their halo 4PCFs contain a large correlated component. When this phase matching is removed by cross-correlating different realization numbers, the signal disappears: the significance falls to $0.0\sigma$ for $\chi^2_{\times}$ and $0.2\sigma$ for $T^2_{\times}$. If the correlated component found in the matched-phased cross-correlation of ODD$_{\mathrm{p}}$ and ODD$_{\mathrm{m}}$ were PV in origin, we should have found a negative-centered distribution and the detection should also be present in the out-of-phase cross-correlation. Instead, we find a positive-centered distribution in the matched-phase cross statistics, whose correlation disappears when conducting the out-of-phase cross statistics. We also investigated the cross-correlation of $\Delta\boldsymbol{\zeta}_{-}$ between Sub~A and Sub~B, whose results also show a zero-significance detection for the halo 4PCF PV signal. Overall, the $\chi^2$ and $T^2$ analyses lead to the same conclusion despite their different treatments of the covariance: There is no PV statistically detectable signal present in the \textsc{Quijote-Odd} halo distribution. The cross-correlation results are summarized in \cref{tab:cross_results}.

\begin{table}
    \centering
    \textbf{Auto-Correlation Results}

    \smallskip
    \begin{tabular}{l S[table-format=+2.1$\sigma$]<{$\sigma$} *{3}{S[table-format=+1.1$\sigma$]<{$\sigma$}}}
        \toprule
        &
        \multicolumn{1}{c}{$\Delta\boldsymbol{\zeta}_{-}$}
        &
        \multicolumn{1}{c}{$\Delta\boldsymbol{\zeta}_{+}$}
        &
        \multicolumn{1}{c}{$\left.\Delta\boldsymbol{\zeta}_{-}\right|_{\mathrm{A}}$}
        &
        \multicolumn{1}{c}{$\left.\Delta\boldsymbol{\zeta}_{-}\right|_{\mathrm{B}}$} \\
        \midrule
        $\chi^2$
        & -39.4
        & -5.5
        & -5.9
        & -4.3 \\
        
        $T^2$
        & -5.3
        & -0.8
        & -4.1
        & -3.4 \\
        \bottomrule
    \end{tabular}
    \caption{Detection significances from the auto-correlation analyses.
    We show the results for $\Delta\boldsymbol{\zeta}_{-}$ and
    $\Delta\boldsymbol{\zeta}_{+}$ measured from the full simulation volume,
    and $\Delta\boldsymbol{\zeta}_{-}$ measured independently in
    Sub A and Sub B. Negative values indicate that the corresponding distribution lies below the Fiducial distribution. For the $T^2$ analysis, we report the results for $N_{\mathrm{eig.}}=100$ eigenmodes. Contrary to expectation, we find $\Delta\boldsymbol{\zeta}_{-}$ to be suppressed with respect to the fiducial distribution. $\Delta\boldsymbol{\zeta}_{+}$ should resemble the fiducial distribution, however it also shows a suppression. The same behavior is recovered in Sub A and Sub B and in both the $\chi^2$ and $T^2$ analyses, although with reduced significance for $T^2$. These results indicate that \cref{eq:zeta_minus_quijote} strongly suppresses the noise shared by the matched realizations, and that the PV signal cannot be recovered for the halo catalogs.}
    \label{tab:auto_results}
\end{table}

\begin{table}
    \centering
    \textbf{Cross-Correlation Results}

    \smallskip
    \begin{tabular}{l S[table-format=2.1$\sigma$]<{$\sigma$} *{2}{S[table-format=1.1$\sigma$]<{$\sigma$}}}
        \toprule
        & 
        \multicolumn{1}{c}{$\boldsymbol{\zeta}_{\mathrm{p}}^{\,i}\times\boldsymbol{\zeta}_{\mathrm{m}}^{\,i}$ }
        & 
        \multicolumn{1}{c}{$\boldsymbol{\zeta}_{\mathrm{p}}^{\,i}\times\boldsymbol{\zeta}_{\mathrm{m}}^{\,j}$ }
        & 
        \multicolumn{1}{c}{$\left.\Delta\boldsymbol{\zeta}_{-}\right|_{\mathrm{A}}
        \times
        \left.\Delta\boldsymbol{\zeta}_{-}\right|_{\mathrm{B}}$} \\
        \midrule
        $\chi^2_{\times}$ 
        & 36.1 
        & 0.0 
        & 0.0 \\
        
        $T^2_{\times}$ 
        & 5.3 
        & 0.2 
        & 0.0 \\
        \bottomrule
    \end{tabular}
    \caption{Detection significances from the cross-correlation analyses.
    The first column corresponds to phase-matched ODD$_{\mathrm{p}}$ and ODD$_{\mathrm{m}}$ realizations, $\boldsymbol{\zeta}_{\mathrm{p}}^{\,i} \times \boldsymbol{\zeta}_{\mathrm{m}}^{\,i}$; 
    the second uses different realization numbers, $\boldsymbol{\zeta}_{\mathrm{p}}^{\,i} \times \boldsymbol{\zeta}_{\mathrm{m}}^{\,j}$ with $i\neq j$; the third cross-correlates $\Delta\boldsymbol{\zeta}_{-}$ measured in 
    Sub~A and Sub~B. For the $T^2_{\times}$ analysis, we report the results for $N_{\mathrm{eig.}}=100$ eigenmodes. A strong positive detection is found when ODD$_{\mathrm{p}}$ and ODD$_{\mathrm{m}}$ realizations with the same phases are cross-correlated, while the signal disappears when different realization numbers are paired. Similarly, the cross-correlation of $\Delta\boldsymbol{\zeta}_{-}$ between Sub~A and Sub~B is consistent with zero. These results show that the large correlation between ODD$_{\mathrm{p}}$ and ODD$_{\mathrm{m}}$ is produced by noise shared by the phase-matched realizations. No PV signal is recovered once this common contribution is removed.}
    \label{tab:cross_results}
\end{table}

\section{Discussion \& Conclusions}
\label{Sec:Discussion}

The main result of this work is that we find no statistically detectable PV signal in the 4PCF of the \textsc{Quijote-Odd} halo catalogs, despite the presence of a PV trispectrum in the initial conditions. This conclusion is supported by several independent tests. The $\Delta\boldsymbol{\zeta}_{-}$ combination, which should enhance the PV contribution, instead produces a distribution strongly smaller than the Fiducial simulations. The $\Delta\boldsymbol{\zeta}_{+}$ distribution, which should resemble the fiducial distribution, is also smaller (although less strongly). In addition, we find a large positive cross-correlation measurement between phase-matched ODD$_{\mathrm{p}}$ and ODD$_{\mathrm{m}}$. This positive correlation disappears when different realization numbers are paired. The same results are found with both the $T^2$ and $\chi^2$ analyses, despite their different covariance treatments. Together, these results indicate that the measured halo 4PCF is dominated by realization-dependent noise, rather than by a PV contribution.

A plausible explanation is that the primordial PV signal does not survive with sufficient strength in the halo distribution. Halos are discrete and biased tracers of the underlying matter field, and the process of halo formation and discrete sampling can reduce or obscure higher-order signatures that are present in the continuous matter distribution. The analysis of \cite{coulton} detects the injected parity-violating trispectrum using the matter distribution, whereas the present work is performed entirely on halo catalogs. Our lack of detection therefore does not imply that the primordial PV signal is absent from the \textsc{Quijote-Odd} simulations themselves, but rather that the signal is not propagated into the PV halo 4PCF considered here.

Although we do not recover the halo-level PV signal, our analysis provides an important validation of how our statistical tests respond to noise. In particular, the strong correlation between phase-matched ODD$_{\mathrm{p}}$ and ODD$_{\mathrm{m}}$ realizations could initially resemble a significant signal (if one forgets that the cross-correlation between ODD$_{\mathrm{p}}$ and ODD$_{\mathrm{m}}$ should be a negative-centered distribution), but the out-of-phase test shows that this correlation is tied to the common Gaussian realization and disappears once that shared component is removed. 

Similarly, the $\text{ODD}_{\mathrm{p}} - \text{ODD}_{\mathrm{m}}$ subtraction strongly suppresses the common fluctuations while leaving a residual noise contribution. The combination of the auto- and cross-correlation tests therefore provides a way to distinguish a parity-odd contribution from a spurious signal generated by noise. This is particularly relevant for observational parity searches, where mismatches between the data and mock catalogs can produce a false detection of PV.

Future work will extend this analysis directly to the \textsc{Quijote} and \textsc{Quijote-Odd} matter distributions. Since the PV trispectrum has previously been detected at the matter level \cite{coulton}, applying the same 4PCF auto- and cross-correlation framework to the matter field will provide a direct test of whether the statistics used here recover the injected PV signal when it is present. Comparing the matter and halo measurements might also help in determining at what stage the parity-odd information is reduced, and what physical process could cause such a loss of information. Such a comparison will be necessary for understanding how primordial parity-violating signatures propagate from the underlying matter field to the discrete tracers used in galaxy surveys.

\section*{Acknowledgments}

WOL acknowledges this material is based upon work supported by the National Science Foundation Graduate Research Fellowship under Grant No.\ DGE-2236414. Any opinions, findings, and conclusions or recommendations expressed in this material are those of the author(s) and do not necessarily reflect the views of the National Science Foundation. ZS acknowledges funding from NASA grant number 80NSSC24M0021. AK was supported as a CITA National Fellow by the Natural Sciences and Engineering Research Council of Canada (NSERC), funding reference \#DIS-2022-568580.  The authors acknowledge UFIT Research Computing for providing computational resources on HiPerGator and support that has contributed to the research results reported in this publication.


\bibliographystyle{JHEP}
\bibliography{references}

\end{document}